# Impact of pedestal density gradient and collisionality on ELM dynamics

Nami Li[1*], X.Q. Xu[1], Y.F. Wang[2], X. Lin[2], N. Yan[2] and G.S. Xu[2]

[1]*Lawrence Livermore National Laboratory, Livermore, CA 94550, USA*

[2]*Institute of Plasma Physics, Chinese Academy of Sciences, Hefei 230031, China*

*[*li55@llnl.gov](mailto:li55@llnl.gov)*

**Abstract**

BOUT++ turbulence simulations are conducted to capture the underlying physics of the small ELM characteristics achieved by increasing separatrix density via controlling strike points from vertical to horizontal divertor plates for three EAST discharges. BOUT ++ linear simulations show that the most unstable modes change from high-n ideal ballooning modes to the intermediate-n peeling-ballooning modes and eventually to peeling-ballooning stable plasmas in the pedestal. Nonlinear simulations show that the fluctuation is saturated at a high level for the lowest separatrix density. The elm size decreases with increasing the separatrix density, until the fraction of this energy lost during the ELM crash becomes less than 1% of the pedestal stored energy, leading to small ELMs. Simulations indicate that small ELMs can be triggered either by the marginally peeling-ballooning instability near the peak pressure gradient position inside pedestal or by a local instability in the pedestal foot with a larger separatrix density gradient. The pedestal collisionality scan for type-I ELMs with steep pedestal density gradient shows that both linear growth rate and elm size decrease with collisionality increasing. While the pedestal collisionality and pedestal density width scan with a weak pedestal density gradient indicate small ELMs can either be triggered by high-n ballooning mode or by low-n peeling mode in low collisionality region 0.04~0.1. The simulations indicate the weaker the linear unstable modes near marginal stability with small linear growth rate, the lower nonlinearly saturated fluctuation intensity and the smaller turbulence spreading from the linear unstable zone to stable zone in the nonlinear saturation phase, leading to small ELMs.

## 1. Introduction

The pulsed heat load due to large ELMs is an existential problem for future devices of ITER size and FPPs because that load would produce unacceptable damage to the divertor plates. On the other hand, the divertor heat flux width for inter-ELMs is too narrow as predicted by the heuristic drift-based (HD) model[1] and an experimental (Eich) scaling[2] due to the suppressed turbulence transport in H-mode plasmas. Therefore, simultaneous control of large ELMs and divertor heat load in H-mode plasma is crucial for steady-state operation of a tokamak fusion reactor[3]. The BOUT++ simulations for small ELMs show the SOL turbulence thermal diffusivity increases due to larger turbulent fluxes ejected from the pedestal into the SOL, leading to the SOL width broadening[4-7]. Recent experiments from current tokamak devices show that small ELM regimes with quasi-continuous exhaust are a promising regime for a reactor with good energy confinement and significantly broadening of divertor heat flux profile. H-mode plasma regime with small/grassy ELMs offers a potential solution for core-edge-integration to future tokamak reactors. However, the identifying the control parameters to access small ELMs regime is still a key physics question. Over the past decades, great efforts have been made on ELM mitigation with different active control methods. The research found that the pedestal electron collisionality is an important parameter that influences the ELM regime. Grassy ELMs have been achieved in highly shaped quasi-double-null (QDN) configuration with high heating power $P_{tot}$~9.5MW and high pedestal collisionality ~4.5 in DIII-D experiments, which shows that the grassy ELM regime is an attractive option for an ELMing scenario in future machines[8]. A high-confinement grassy-ELM H-mode regime has been achieved since 2016 in EAST with high pedestal collisionality ~1[9]. On DIII-D, a grassy ELM regime with low pedestal collisionality ~0.15 is achieved with and without Resonant Magnetic Perturbations (RMP) control, showing a consistent divertor heat flux width broadening and peak amplitude reduction with BOUT++ prediction[10,11]. High confinement, and power exhaust compatible H-mode regimes with small ELMs have been achieved in TCV and ASDEX-Upgrade (AUG) with high separatrix density[12]. A new H-mode regime with small ELMs and high confinement in the JET-ILW has also been obtained at the low edge collisionality values expected in ITER both in D-D and D-T plasmas, by operation with low or no-gas and pellet injection, named Baseline Small ELM (BSE) regimes[13].

Recent EAST experiments show for the first time that the small/grassy ELM regime can be achieved via controlling the pedestal density profiles by changing striking point from vertical to horizontal target plates on the new lower tungsten divertor[14]. The merely changes of the striking point position leads to significantly flattening pedestal density profiles, increased ELM frequencies and reduced ELM amplitudes. This discovery provides an alternative knob for ELM control and demonstrates that a low pedestal density gradient is a key for access to small-ELM regimes and a wide pedestal can lead to an ELM suppression. Small ELMs have been achieved on different machines with different separartrix density and collisinalit[8-13]. Therefore, it is very important to understand the underlying physics how pedestal density gradient and collisonality effect on ELMs dynamics and provide control knobs for the access to small ELM regime for ITER, as ITER is planning to operate in a regime with low pedestal collisionality, low pedestal density gradient and high separatrix collisionality[15].

To investigate the impact of the pedestal density gradient and collisionality on the ELMs dynamics, we first perform BOUT++ turbulence simulations for EAST experimental discharges with different pedestal density profiles achieved via controlling strike points from vertical to horizontal divertor plates. In order to project from the current EAST tokamak experiments to ITER relevant parameters, we then conduct BOUT++ simulations for the pedestal collisionality scan with a fixed pedestal pressure by decreasing pedestal density and increasing pedestal temperature. Furthermore, we carry out BOUT++ simulations for a scan of the pedestal density gradient with a fixed temperature profile. The paper is organized as follows: Section 2 provides a description of experiment parameters and simulation settings. The simulation results for three EAST discharges with different density profiles are shown in Section 3, including (1) linear MHD stability analysis, (2) characterization ELM nonlinear dynamics for three EAST discharges with different pedestal density profiles. Scans for the pedestal collisionality and the density gradient or width are presented in Sec.4 and finally, a summary of the results is given in Sec. 5.

## 2. Magnetic equilibria, plasma profiles and simulation settings

A set of three discharges with different pedestal density profiles obtained via controlling strike points from vertical to horizontal divertor plates are used for simulations with EAST lower single null (LSN) divertor configuration. The main plasma global parameters for these three discharges are similar except the lower divertor strike point position which leads to different separatrix density, as shown in Table 1 with shot numbers #103751, #103745 and #103748. Large ELMs are observed with the strike point located on the vertical target for EAST shot #103751 with ELM frequency $f_{ELM}$ ~120. While small ELMs and even smaller ELMs are obtained with the strike point on the horizontal target for shot #103745 with $f_{ELM}$ ~300 and #103748 with $f_{ELM}$ ~500, respectively, which are in a typical frequency range of small grassy ELMs. The energy confinement is maintained in these three cases with $H_{98y2}$ ~1.0. The kinetic equilibria are reconstructed using the EFIT code[16-18] with the constraints of experimentally measured total pressure profile and flux surface averaged toroidal current density profile dominated by bootstrap current $j_{BS}$ in the pedestal region. The radial equilibrium profiles of pressure and current density for these three shots are shown in Fig. 1(a) and (b). The simulations include the plasma edge, the SOL region and private flux region. The simulation domain is shown in Fig. 2, which ranges from normalized poloidal flux $\psi=0.75$ to $\psi=1.05$, where $\psi=1.0$ is the magnetic separatrix as shown by the red curve. The spatial resolution of the grid generated from the equilibrium file (EFIT g-file) is 260 radial grid points and 64 poloidal grid points. In the poloidal direction, there are 4 grid points for private flux region of each divertor and 56 grid points for the main plasma region above the x-point for LSN divertor configuration. In the toroidal direction, we only simulate one fifth of the torus for the nonlinear simulations for simplicity.

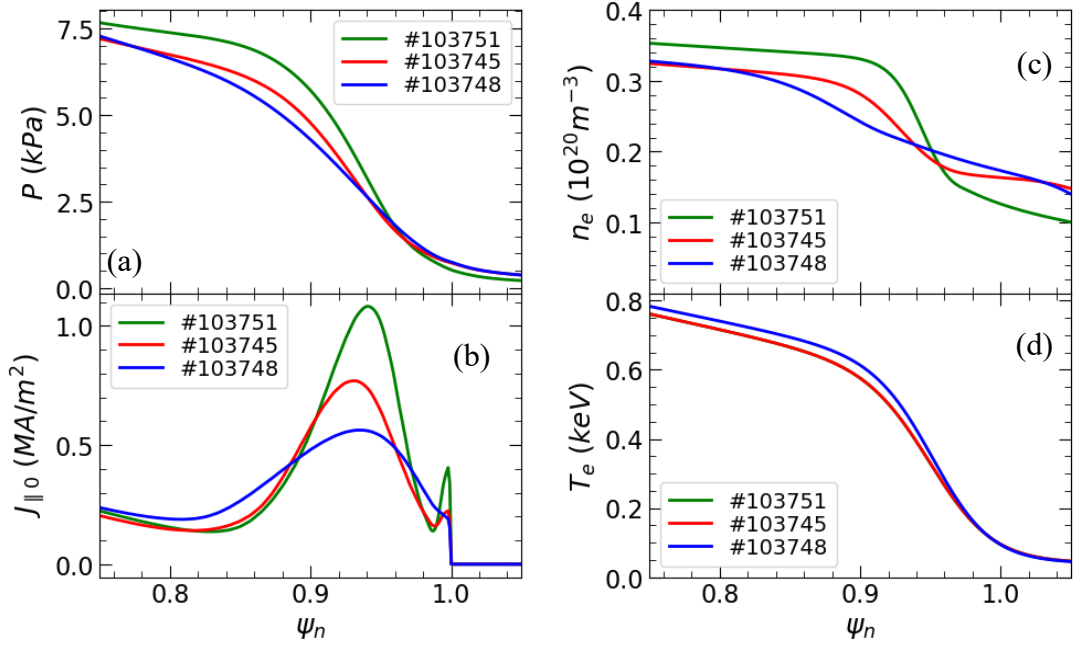

Fig. 1 Plasma profiles for BOUT++ simulations with three EAST discharges #103751(green), #103745(red) and #103748(blue): (a) the pressure profiles; (b) current density profiles; (c) electron density and (d) electron temperature profiles.

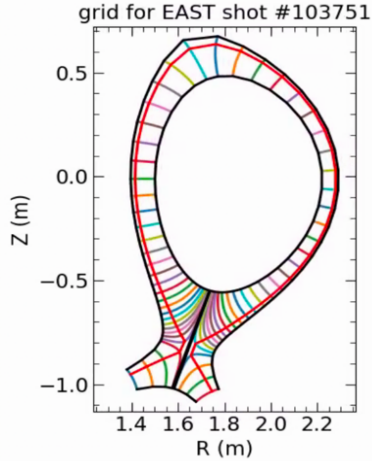

Fig. 2 Magnetic geometry and grid for BOUT++ simulation with shot #103751.

The initial plasma profiles used in BOUT++ simulations are taken from fits of a modified hyperbolic tanh function[19] to experimental data measured by microwave reflectometry for density $n_e$ and Thomson scattering for temperature $T_e$, mapped onto a radial coordinate of normalized poloidal magnetic flux with the SOL region. The electron temperature profiles for these three shots are similar as shown in Fig. 1(d), while the density profiles exhibit a dramatic change as shown in Fig. 1(c). The change in pedestal pressure profile is mainly induced by the density profile change. With the strike point changing from the vertical target to the horizontal target, the separatrix density and the density ratio between the separatrix and pedestal keep increasing, while the pedestal density gradient decreases correspondingly. Thus, the pedestal pressure gradient and bootstrap current density are significantly reduced when the strike point changing from the vertical target to the horizontal target.

In this work, BOUT++ two-fluid six-field turbulence simulations are conducted to understand the ELM dynamics, which evolve the perturbations of vorticity $\varpi$, ion density $n_i$, ion and electron temperature $T_i$ and $T_e$, ion parallel velocity $V_{\|i}$ and parallel magnetic vector potential $A_\|$. In the model, both ideal MHD and non-ideal physics effects are included, such as Peeling–Ballooning instability, ion diamagnetic effects, resistivity, hyper-resistivity, the first order ion finite Larmor radius effect due to the gyro-viscous stress tensor, parallel thermal conductions, Hall effects, toroidal compressibility, electron–ion friction, etc. The simple form of the parallel viscosity $\mu_{\|i}\nabla^2_{\|0}\varpi$ is used in the vorticity equation while the perpendicular viscosity is neglected. The flux-limiting parallel thermal conductivities are used to make up the kinetic correction of parallel transport from collisionless to collisional regimes. The perpendicular classical diffusivities are neglected in the simulations. Sheath boundary conditions are imposed on the divertor targets. Neumann boundary condition is applied on inner radial boundary while Dirichlet boundary condition for outer radial boundary. For the core region, twist-shift periodic boundary condition is set in y direction and periodic boundary condition is used in toroidal direction. More detailed settings can be found in previous papers[20,21]. The simulation settings are same for these three discharges.

Table 1 Parameters of three EAST discharges

| shot | $I_p$ (kA) | $\beta_p$ | $\beta_N$ | $q_{95}$ | $H_{98,y2}$ | $\nu^*_{ped}$ | $n_{e,sep}(10^{19}m^{-3})$ | $n_{e,sep}/n_{e,ped}$ |
|---|---|---|---|---|---|---|---|---|
| #103751 | 452 | 1.35 | 1.26 | 5.74 | ~1.0 | 1.46 | 1.26 | 0.38 |
| #103745 | 451 | 1.31 | 1.24 | 5.81 | ~1.0 | 1.08 | 1.63 | 0.53 |
| #103748 | 449 | 1.32 | 1.24 | 5.89 | ~1.0 | 0.78 | 1.73 | 0.55 |

## 3. The characteristics of ELM dynamics for three EAST discharges with different pedestal density profiles

To capture the characteristics of ELM dynamics for three EAST discharges (#103751, #103745 and #103748) with different pedestal density profiles achieved via controlling strike points from vertical to horizontal divertor plates, in this section, both linear and nonlinear simulations are performed using BOUT++ two-fluid six-field turbulence code. To investigate the SOL density gradient effect on the ELM dynamics with a stable pedestal, the comparison of two nonlinear simulations with different SOL density gradient profiles are shown in section 3.2.1.

**3.1 Linear MHD stability analysis**

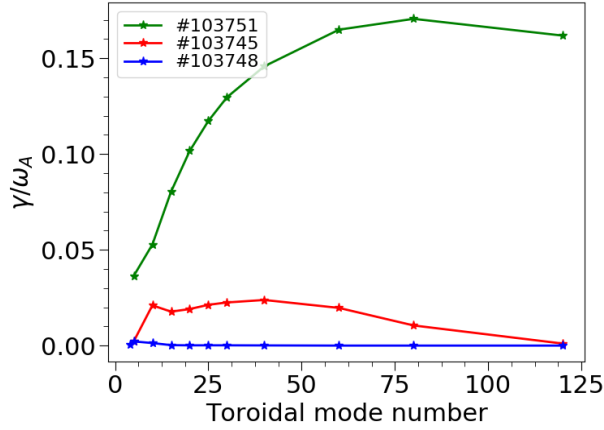

Fig. 3 Linear growth rate from the linear toroidal mode number scan for shot #103751 (green), #103745 (red) and #103748 (blue).

Firstly, in order to understand the dominant modes for these three discharges, we focus on linear simulations by turning off the nonlinear terms in the equations of six-field two-fluid turbulence code. The linear simulations start from small random perturbations. Fig. 3 shows the toroidal mode number spectrum of the linear growth rate at the location of the peak gradient of pedestal pressure. The horizontal axis is the toroidal mode number, and the vertical axis is the linear mode growth rate normalized by the Alfven frequency at the magnetic axis, $\omega_A = \frac{B_0}{R_0\sqrt{\mu_0 n m_i}}$. BOUT++ linear simulations for ideal MHD instability with ion diamagnetic stabilization effect show that the linear growth rate dramatically decreases and the dominated mode shifts from the high-n to intermediate-n modes due to the pedestal pressure gradient dramatically reduced, as separatrix density increases and pedestal density gradient decreases. For shot #103751, the most unstable toroidal mode number is at n~80 with characteristics of ballooning mode driven by the steep pedestal pressure gradient, as shown by the green curve in Fig. 3. A large ELM will be triggered in the nonlinear simulations driven by ballooning mode with large linear growth rate. As the pedestal density gradient decreases, the most unstable toroidal mode number shifts from high-n to intermediate-n (n~40) with characteristics of peeling-ballooning (P-B) mode for shot#103745 and the linear growth rate decreases dramatically due to significantly reduced pedestal pressure gradient and current density, as shown by the red curve in Fig. 3. Small ELMs are triggered in the nonlinear simulations with marginally unstable peeling-ballooning mode. For these two cases, since the pedestals are unstable to peeling–ballooning modes, perturbations grow up around the location of the peak gradient of pedestal pressure at the outer midplane (OMP). As the separatrix density and pedestal density width further increase, the linear growth rate further decreases, and the peeling-ballooning modes are stable in the pedestal region for shot #103748 as shown by the blue curve in Fig. 3. However, a local mode grows up near the separatrix driven by the large density gradient near the separatrix with high separatrix density which triggers the small ELM in the nonlinear simulations, even though the peeling-ballooning mode is stable in the pedestal. More details about the ELM nonlinear analysis will be presented in next section 3.2.

### 3.2 Characteristics of ELM nonlinear dynamics for three EAST discharges

BOUT++ two-fluid six-field nonlinear turbulence simulations are conducted to capture the physics of the ELM dynamics with different pedestal density for shot #103751, #103745 and #103748. The same simulation settings for these three discharges are used in the nonlinear simulations, as in the linear simulations in Sec.3.1 except that all nonlinear terms are turned on. The radial simulation domain covers from normalized poloidal flux $\psi_N = 0.75$ to $\psi_N = 1.05$ for three discharges, crossing the separatrix.

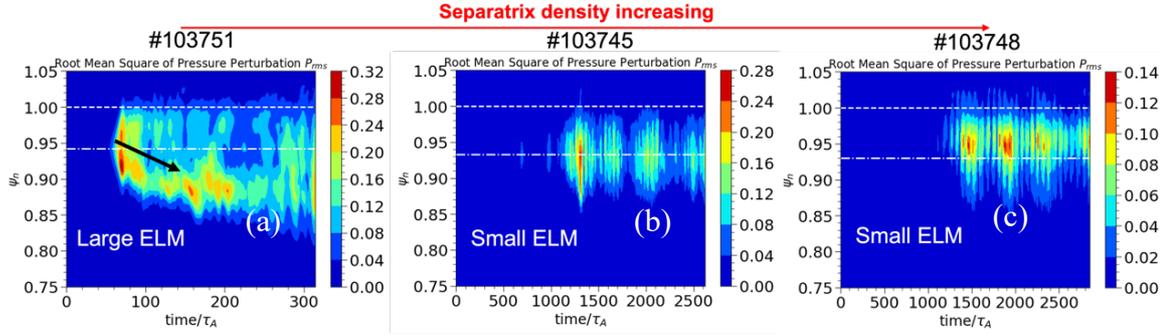

Fig. 4 Spatial-temporal evolution of root mean square (RMS) of pressure perturbation at the outer midplane (OMP) for three discharges #103751(a), #103745(b) and #103748(c). The white dashed line is at separatrix, and the dashed-dot line is at the location of peak gradient of pressure.

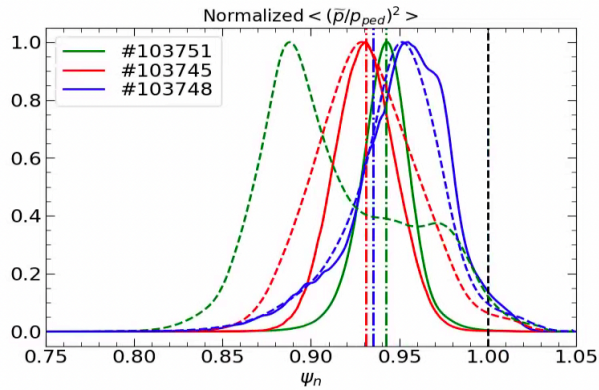

Fig. 5 The mode structure of normalized pressure fluctuation intensity at OMP in linear (solid curve) and nonlinear phase (dashed curve). The green curve is for shot #103751. The red curve is for shot #103745 and the blue curve is for shot #103748.

The spatial-temporal evolution of the pressure perturbation further illustrates the generation of the turbulence. Fig. 4 shows the contour plot of the root-mean-square (RMS) value of pressure perturbation at the outer midplane vs time and radius for three discharges #103751(a), #103745(b) and #103748(c). The pressure perturbation is normalized by equilibrium pressure at the pedestal top, the time is normalized by Alfven time and radius is a normalized poloidal flux. For shot #103751, with low separatrix density and steep pedestal density gradient, the pedestal is unstable to ballooning modes and the linear phase lasts in a short period of time $\sim 80\tau_A$ with a large linear growth rate, which is consistent with linear simulation in Fig. 3. The fluctuation generates at the location of peak pressure gradient in the linear stage and a large ELM is triggered at the early nonlinear stage driven by the ideal ballooning mode. The fluctuation keeps spreading inward to the pedestal top after the ELM triggered and finally saturates at a high level around 15% at the pedestal top as shown by the Fig. 4(a). As the separatrix density increases, for shot #103745, the pedestal density and pressure profiles appear to be broadened. The pedestal density gradient is significantly reduced with a higher density ratio between the pedestal bottom (separatrix) and top $n_{e,sep}/n_{e,ped} \sim 0.53$. The pedestal pressure gradient and bootstrap current density is significantly reduced correspondingly. The pedestal is marginally unstable for ideal peeling-ballooning mode. The linear phase lasts in a long period of time $\sim 1200\tau_A$ with a small linear growth rate. The fluctuation generates at the location of peak pressure gradient in the linear stage which is same as the shot #103751 with a large ELM. A small ELM is triggered at the early nonlinear stage and saturates at a low level around 9.6% at the position of peak pressure gradient, as shown by Fig. 4(b). When the separatrix density further increases, for shot #103748, an even flatter density profile is obtained with a high density ratio $n_{e,sep}/n_{e,ped} \sim 0.55$. The pedestal is stable for ideal peeling-ballooning mode in the linear simulation. However, due to a local large density gradient near the separatrix in the bottom of pedestal, the nonlinear simulation show that strong turbulence can be generated at the bottom of pedestal after a long period of linear growing phase which is lasts $\sim 1300\tau_A$ and can be inward spread into the pedestal region. Finally, the fluctuation saturates at the pedestal bottom with a relative lower level $\sim 6.6\%$ in comparison with shot #103745 at 9.6% and shot #103751 at 15%, as shown by Fig. 4(c). A small ELM can also be triggered driven by density gradient near the separatrix even though the ideal P-B mode is stable in the pedestal.

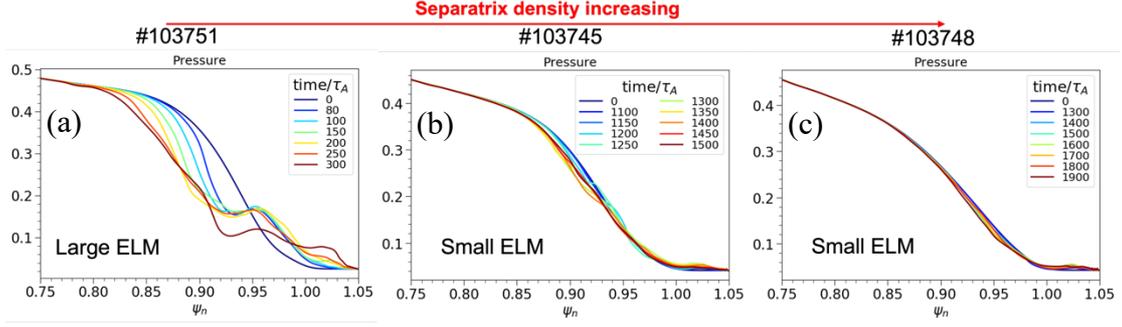

Fig. 6 The pressure profiles at different time slices for these three discharges #103751(a), #103745(b) and #103748(c).

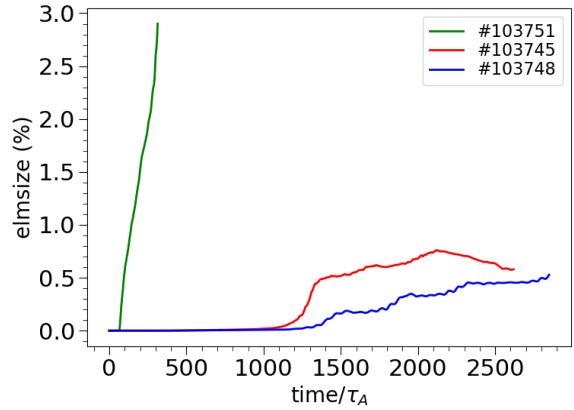

Fig. 7 Time evolution of 3D relative ELM size. The green curve is for shot #103751. The red curve is for shot #103745 and the blue curve is for shot #103748. The ELM size is defined as elmsize $= \Delta W_{ped}/W_{ped}$, where $\Delta W_{ped} = \int_{R_{in}}^{R_{out}} \oint dRd\theta (P_0 - \langle P \rangle_\zeta)$ is the ELM energy loss and $W_{ped} = \frac{3}{2} P_{ped} V$ is the stored energy.

To further compare the mode structure for different type ELMs, the radial mode structures of normalized pressure perturbation intensity at outer midplane in the linear and nonlinear stage are shown in Fig. 5. For large ELM (shot #103751), in the linear stage, the mode grows up at the pressure peak gradient locations as shown by the solid green curve in Fig. 5 and the fluctuations localize in the pedestal region. While in the nonlinear stage, as shown by the dashed green curve in Fig. 5, a large turbulence spreading is observed. The peak location of the mode shifts inward in the nonlinear saturated phase. The fluctuations spread inward and outward on both sides of the pedestal region, reaching to the pedestal top region and the SOL, indicating a large collapse of pedestal pressure as shown in Fig. 6(a), a characteristic of an ELM burst. The time evolution of radial pressure profile demonstrates the flattening process during the ELM burst. After the ELM is triggered, the pressure profile keeps dropping inside the separatrix while rising in the SOL, which means that energy is continuously transferred from the edge region to the SOL. Therefore, the ELM size keeps increasing, which is a large ELM with ELM size larger than 1%, as shown by the green curve in Fig. 7. Here, the ELM size is defined as elmsize = $\Delta W_{ped}/W_{ped}$, where $\Delta W_{ped} = \int_{R_{in}}^{R_{out}} \oint dRd\theta (P_0 - \langle P \rangle_\zeta)$ is the ELM energy loss and $W_{ped} = \frac{3}{2} P_{ped} V$ is the stored energy. As the separatrix density increases, the nonlinear simulation for the small ELM with shot #103745 show that, in the linear stage, the mode also grows up at the peak pressure gradient locations as shown by the solid red curve in Fig. 5. The mode structure in the linear phase is similar with shot #103751. While, in the nonlinear phase, the mode structure is mostly unchanged and localized in the pedestal steep-pressure gradient region as shown by the dashed red curve in Fig. 5, and the flattening of pressure profile is localized only in a small radial area around the middle of the pedestal as shown by Fig. 6(b), where at the peeling-ballooning mode (PBM) peaks. The pedestal falls into the linear stable zone just shortly after the initial ELM crash. Thus, both instability growth and pedestal collapse stop, and the ELM size remains at less than 1% (red curve in Fig. 7), leading to a small ELM. As the separatrix density further increases, the small ELM is also triggered even though the ideal PBM is stable in the pedestal for shot #103748. In the nonlinear simulation, we found that the local mode grows up at the bottom of pressure pedestal (near separatrix) in the linear stage due to a large density gradient driven near the separatrix. A strong fluctuation is generated near the separatrix, and the pressure profile marginally collapses inside the separatrix, which then spread inward into the pedestal as show by the Fig. 6(c). Finally, the fluctuation is localized in the pedestal bottom region (outside the steep-pressure gradient region) as shown by the blue curves in Fig. 5 and the ELM size also remains less than 1% (blue curve in Fig. 7), leading to a small ELM. Overall, BOUT++ simulations results show a consistent trend with the experimental observations that the ELM size decreases into small ELM regime with increasing separatrix density for shot # 103745 and #103748. From the experiment, the fraction of the energy lost during the ELM crash is around 3.25% of the pedestal stored energy based on the calculation by EFIT for shot #103751, which is comparable with the BOUT++ nonlinear simulation result. While for shots #103745 and # 103738, due to the limited resolution of experimental diagnostics, it is hard to get the actual energy loss for such small ELM. However, the diagnostic of edge density from POINT show that the amplitude of density fluctuation decreases from large ELM (shot#103751) to small ELM (shot # 103745 and #103748).

**3.2.1 Effect of separatrix density gradient on the ELM dynamics**

Based on the nonlinear simulation for shot #103748, we can find that the small ELM can be triggered by the local instability with large density gradient near the separatrix. To investigate the separatrix density gradient effect on the ELM dynamics, two nonlinear simulations are conducted with same equilibrium and same settings except the density profiles in the SOL for shot #103748, as shown in Fig. 8(a). The red curve shows the case with a flat density profile in the SOL. The density gradient near the separatrix is much small. The blue curve shows the case with a steep density gradient near the separatrix. The nonlinear simulation with a flat SOL density profile shows that a weak density fluctuation is generated in the pedestal region in the linear phase and finally saturates at a low level ~0.6%, as shown by Fig. 8(c). There is no ELM triggered because the ideal peeling-ballooning mode in the pedestal is stable and there is no local insatiably driven due to the flat density near the separatrix. Therefore, from the linear phase to the nonlinear phase, there is no sudden collapse of the pressure as shown in Fig. 8(b), where the black curve overlaps with dashed red curve. Here the black curve is the initial pressure profile, and the dashed red curve is the pressure profile in the nonlinear phase with flat SOL density. While with a steep density gradient near the separatrix, a local mode grows up and a strong density fluctuation is generated near separatrix first, which modifies the pressure profile near the separatrix and the pressure gradient at the pedestal bottom increases marginally, leading to the ballooning mode unstable at the pedestal bottom. A strong density fluctuation is generated at the bottom of pedestal in the later linear phase and then spread both inward and outward. Finally, the peak density fluctuation oscillates between the position of peak pressure gradient and the separatrix. A small collapse of pressure occurs at the pedestal bottom, a characteristic of a small ELM burst, as shown by the dashed blue curve in Fig. 8(b).

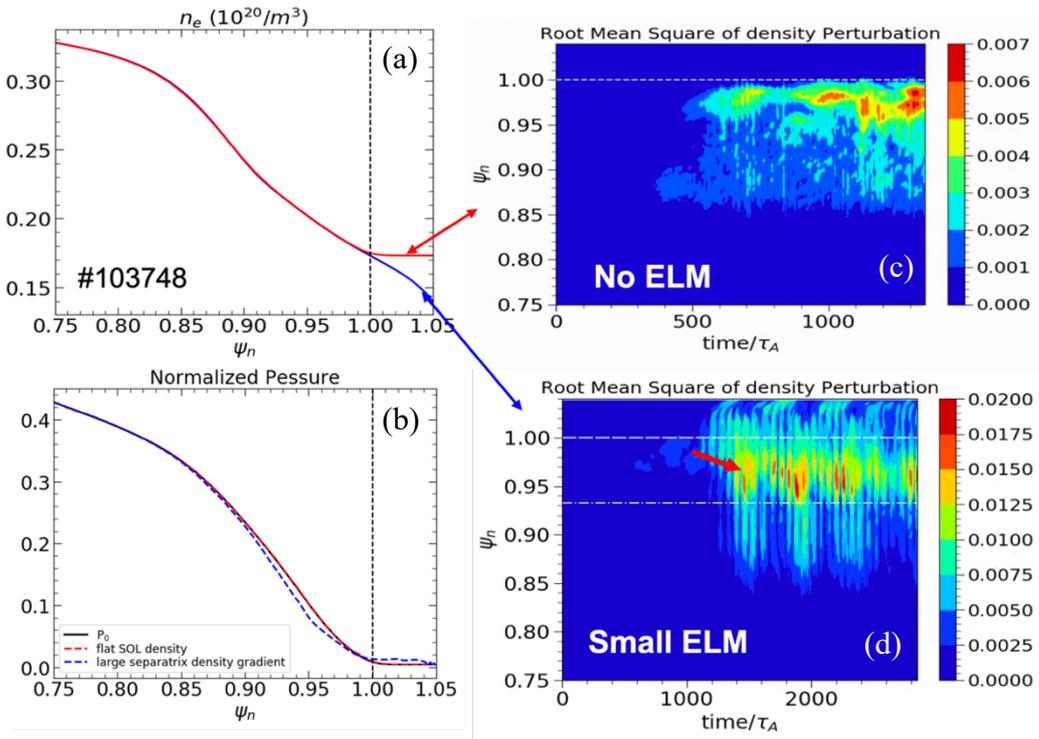

Fig. 8 (a) Initial denisity profiles and (b) pressure profiles. (c) Spatial-temporal evolution of RMS of density perturbation at the OMP with flat SOL density and (d) with large separatrix density gradient for shot #103748.

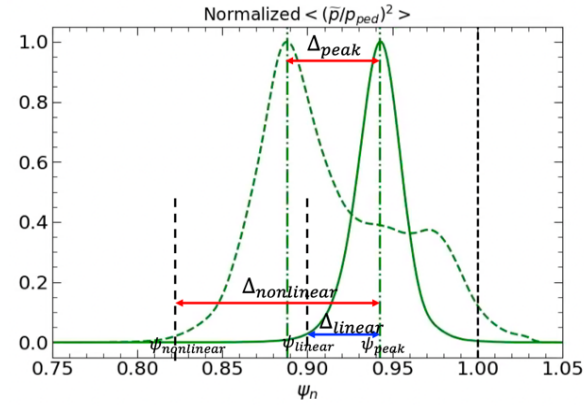

Fig. 9 Radial profiles of the normalized pressure fluctuation intensity in the linear (solid curve) and nonlinear phase (dashed curve) for the shot #103751.

Overall, the nonlinear simulations for these three discharges with different pedestal density profiles indicate the small ELM can be triggered either with the marginally unstable ideal peeling-ballooning mode near the peak position of the pressure gradient (#103745) or by local marginal ballooning instability near separatrix with larger density gradient (#103748). The underlying physics which keeps the ELM small or large strongly depends on the ELM affected area as shown in Fig. 5 by the mode structure of normalized pressure fluctuation intensity at OMP from linear (solid curve) to nonlinear phase (dashed curve). The ELM affected area is strongly impacted by the inward avalanche and turbulence spreading from the linear unstable zone to stable zone in the nonlinear saturation phase, which includes the front propagation and penetration, as shown in Fig. 9. Here we define the front propagation as the convective movement of peak intensity position $\Delta_{peak}$ and the diffusive penetration as the intensity radial profile broadening. To explore the correlation between pressure fluctuation intensity and avalanche/turbulence spreading, the time and binormal averaged pressure fluctuation intensity at OMP $I_p$ in the nonlinear saturated phase is calculated, $I_p = \langle \left(\frac{\tilde{p}}{P_{ped}}\right)^2 \rangle$. To define the extent of avalanche/turbulence spreading, we introduce the ELM affected area factor as $\gamma_s = \frac{\Delta_{nonlinear}}{\Delta_{linear}}$ as shown by the Fig. 9. Here, the penetration depth $\Delta$ is defined by where the front foot starts to decrease to $10^{-2}$ of the peak fluctuation intensity level. The $\Delta_{linear} = \psi_{peak} - \psi_{linear}$ is the linear mode depth and the $\Delta_{nonlinear} = \psi_{peak} - \psi_{nonlinear}$ is the penetration depth in the nonlinear phase. For the large ELM (#103751), the linear mode is very unstable with large linear growth rate at the peak gradient of pedestal pressure and the pressure fluctuation intensity at the onset of nonlinear phase is strong. After the initial ELM is triggered, the original pedestal profile collapses, which results in the profile steepening inward near the pedestal top. The pedestal top gets into the linear unstable zone, which leads to a 2$^{nd}$ collapses. These processes continue to generate multiple collapses inward until the crashing stops, leading to the front propagation. A strong inward avalanche with the front propagation occurs for the type-I ELM. The multiple profiles collapse from linear unstable zone near original pedestal peak gradient position into stable zone in the pedestal top region, which tap more pedestal plasma stored energy (associated with the inhomogeneities in the plasma current, pressure and magnetic field) into fluctuation energy, leading to large ELMs. While for small ELMs, the pedestal is near linear instability threshold and pressure fluctuation intensity at the onset of nonlinear phase is much weaker, the fluctuation is limited near the peak gradient position of the pedestal with only diffusive penetration, but without the front propagation. The ELM size is correlated with either the avalanche process for type-I ELMs or turbulence spreading process for small ELMs. From the large ELM to the small ELM, the ELM affected area factor decreases from $\gamma_s$ ~2.56 (#103751) to $\gamma_s$ ~1.5 (#103745) and $\gamma_s$ ~1.06 (# 103748) as the saturated pressure fluctuation intensity decreases from 2.67% (shot #103751) to 1% (#103745) and 0.5% (#103748). Therefore, we observe that the weaker the linear unstable modes near marginal stability, the lower nonlinearly saturated fluctuation intensity and the smaller the turbulence spreading, leading to small ELMs.

## 4. The impact of collisionality and pedestal density gradient on the ELM dynamics

Recent studies indicate that collisionality and pedestal density gradient play an important role for access to small-ELM regime. As we mentioned in the introduction, the small/grassy ELMs regime have been achieved from experiment in different machines with different pedestal collisionality $v_{ped}^*$: such as 1) grassy ELMs regime on EAST with high $v_{ped}^*$~1[9]; 2) grassy ELMs regime on D-IIID with low $v_{ped}^*$~0.15[10]; 3) Quasi-Continuous Exhaust (QCE) regime on ASDEX-U and TCV with high separatrix density/collisionality[12]; 4) small ELMs regime with low gas & pellets injection on JET-ILW[13] in both D-D and D-T plasma (pedestal collisionality is low and close to ITER). However, it is still an open question whether there is still the access to small/grassy ELM regime for ITER, which will have a very low pedestal collisionality $v_{ped}^*$~0.01 with low pedestal density gradient $\nabla n$. Therefore, it is important to investigate the impact of pedestal collisonality and pedestal density gradient on the ELM dynamics. The pedestal collisionality scan is performed in section 4.1 with a set of equilibria based on shot #103751 for large ELMs with large pedestal density gradient and shot #103748 for small ELMs with weak pedestal density gradient, respectively. The pedestal density gradient scan is performed in section 4.2.

## 4.1 Collisionality scan for type-I ELM and small ELM

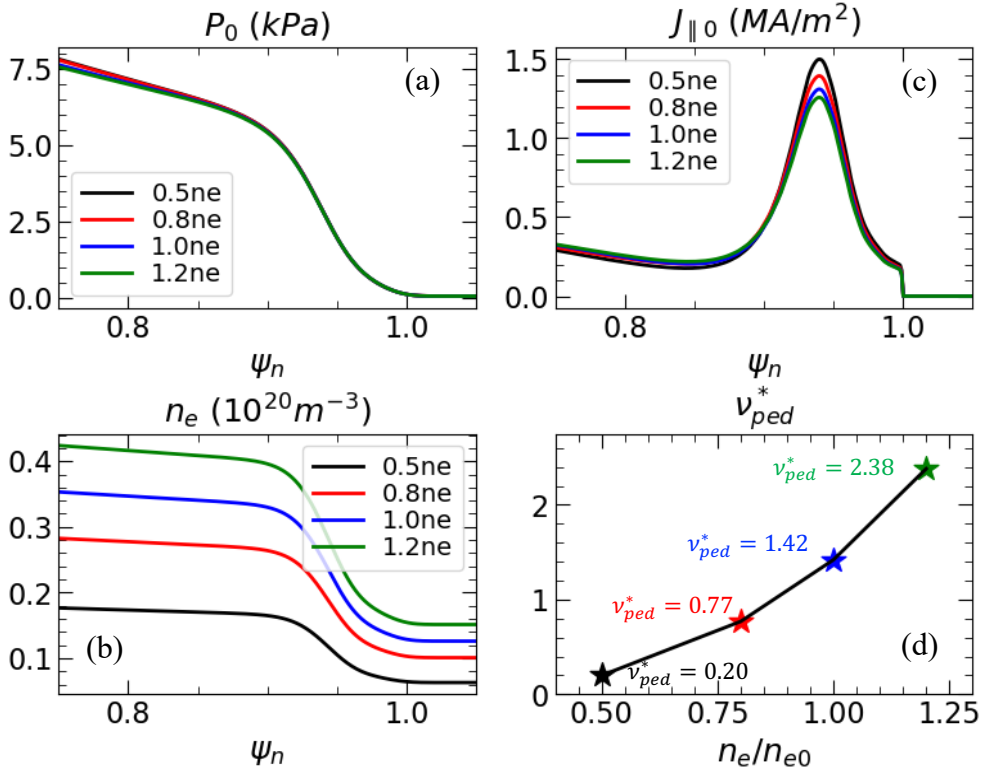

Fig. 10 Plasma profiles with collisionality scan for shot #103751. (a) pressure profiles, (b) density profiles and (c) current density profiles with collisionality scan. (d) normalized pedestal electron collisionality with different density.

First, starting from the type-I ELM, the pedestal collisionality scan is performed with fixed pressure profile for EAST discharge #103751. Based on the experimental profiles, the density increases by multiplying a factor 0.5x, 0.8x, 1.0x and 1.2x as shown in Fig. 10 (b), while the temperature decreases accordingly by dividing a same factor 0.5x, 0.8x, 1.0x and 1.2x, the pedestal collisionality varies by an order of magnititude. A series of the equilibria are reconstructed using the kinetic EFIT code with these modified density and temperature profiles. In order to understand the impact of the collisionality on the pedestal instability and therefore on the ELM, initial flat plasma profiles in the SOL are assumed in the following simulations, which eliminate possible local SOL instabilities. The initial profiles are shown in Fig. 10. The collisionality increases as the density increases and temperature decreases as shown in Fig. 10(d), and thus the bootstrap current density reduces accordingly as shown by Fig. 10(c). Table 2 shows the pedestal collisionality calculated by the formula:

$$v_{ped}^* = 6.921 \times 10^{-18} \frac{Rqn_e Z_{eff} \ln \Lambda_e}{T_e^2 \epsilon^{3/2}} \qquad (1).$$

Table 2. the pedestal collisionality with density scan for shot #103751.

| $n_e$ | 0.5x | 0.8x | 1.0x | 1.2x |
|---|---|---|---|---|
| $v_{ped}^*$ | 0.20 | 0.77 | 1.42 | 2.38 |

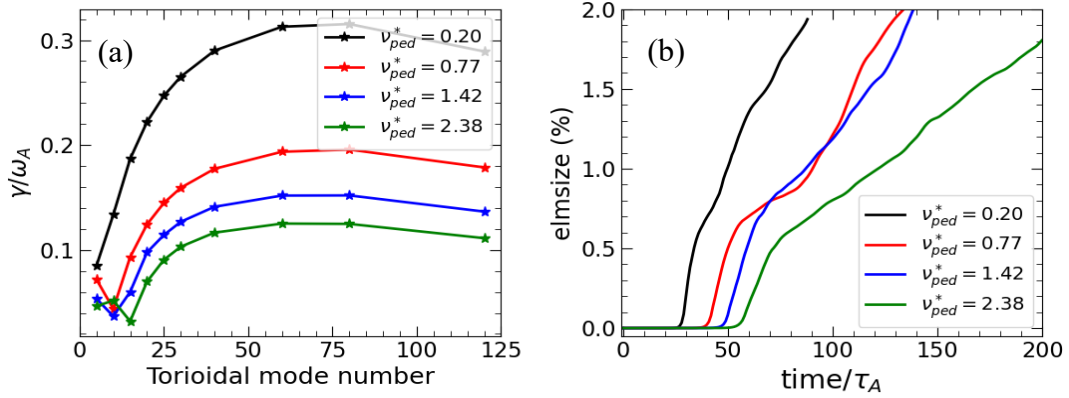

Fig. 11 (a) Linear growth rate from the linear toroidal mode number scan and (b) time evolution of 3D relative ELM size with collisionality scan for shot #103751.

Both linear and nonlinear simulations are performed using BOUT++ turbulence six-field two-fluid code. Fig. 11(a) shows the linear growth rate vs toroidal mode number n and (b) shows the time history of the ELM energy loss fraction for the density/collisionality scan. The pedestal is unstable for ideal ballooning mode with high-n (n~80) due to the steep pedestal pressure gradient. As collisionality increases, linear growth rate decreases while the dominant mode remains in the high-n range even though the current marginally reduces. The general trend is that the plasma with the lower density (lower collisionality) has larger linear growth rate, faster front propagation and deeper penetration, reaching inner boundary sooner and leading to a sharply increased energy loss, as shown by the black curve in Fig. 11. The simulation results of the ELM size reduction with increasing pedestal plasma collisionality for EAST type-I ELM show a good agreement with the multi-ITPA experimental database[22] and previous BOUT++ simulation results using a set of circular cross-section toroidal equilibria[23]. It is worth noting that even though the ELM size decreases with the collisionality increasing, but a clear front propagation is also observed for the highest collisionality $v_{ped}^* = 2.38$, it is still characterized as a small type-I ELM with clear avalanche spreading.

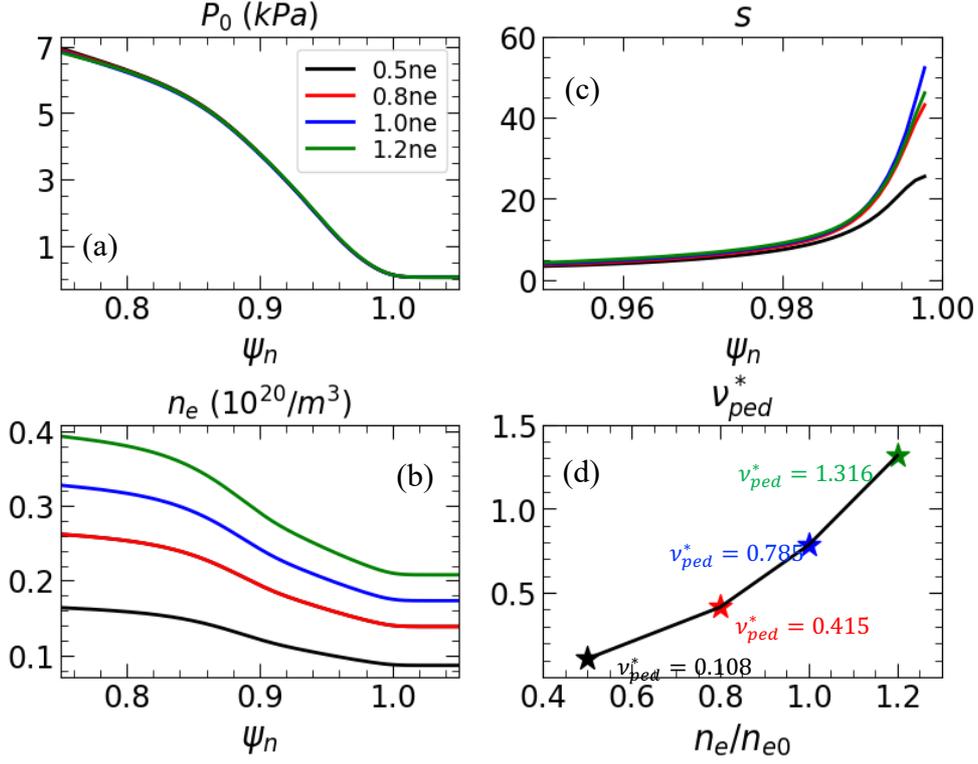

Fig. 12 Plasma profiles with collisionality scan for shot #103748 with flat SOL density profile. (a) pressure profiles, (b) density profiles and (c) magnetic shear with collisionality scan. (d) normalized pedestal electron collisionality with density scan.

Table 3. the pedestal collisionality with density scan for shot #103748 with flat SOL density profiles.

| $n_e$ | 0.5x | 0.8x | 1.0x | 1.2x |
|---|---|---|---|---|
| $\nu_{ped}^*$ | 0.108 | 0.415 | 0.785 | 1.316 |
| | | | | |
| $n_e$ | 0.5x | 0.4x | 0.3x | 0.2x |
| $\nu_{ped}^*$ | 0.108 | 0.069 | 0.039 | 0.017 |

Both the simulations and experiments for EAST show that a low pedestal density gradient is a key for access to small-ELM regimes. As we know, ITER will operate in the low pedestal collisionality regime with low pedestal density gradient[15]. Here is a question: will ITER be able to access the small ELM regime with wide pedestal density but low pedestal collisionality? Based on EAST shot #103748 for small ELMs, additional BOUT++ simulations for a wider collisionality scan have been performed as follows by decreasing density and increasing temperature with fixed pressure, including the low edge collisionality values expected in ITER. As we show in the section 3, a small ELM will be triggered driven by local instability with a large density gradient near the separatrix. While the ELM will disappear with a flat SOL density profile. Therefore, to investigate the impact of the pedestal collisionality on the pedestal MHD instability, a flat SOL density profile is assumed to ignore the local instability near the separatrix or in the SOL. Fig. 12 shows the plasma profiles for the collisionality scan with flat SOL density profiles. The pressure profiles are shown in Fig. 12(a) and density profile shifts by multiplying a factor 0.5x, 0.8x, 1.0x, and 1.2x as shown in Fig. 12(b). The collisionality decreases as the density decreases for one order of magnitude, as shown in Fig. 12(d), and thus the magnetic shear reduces accordingly as shown by Fig. 12(c). Table 3 shows the pedestal collisionality calculated by the formula (1) for the collisionality scan of shot #103748.

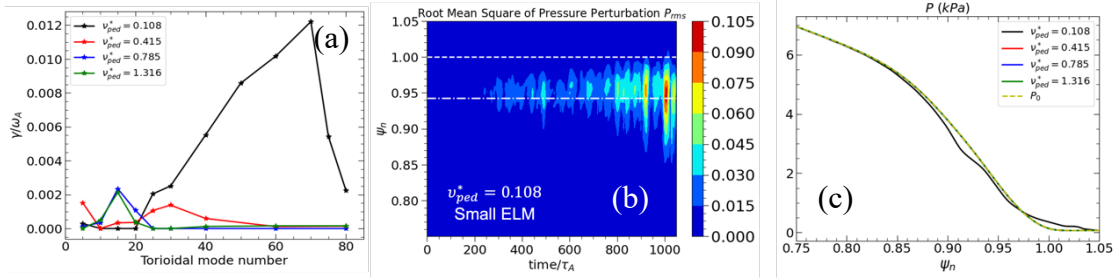

Fig. 13 (a) Linear growth rate from the linear toroidal mode number scan; (b) Spatial-temporal evolution of RMS of pressure perturbation at the OMP with lower collisionality $v_{ped}^* = 0.108$ and (c) pressure profiles at nonlinear phase with different collisionality for shot #103748 with flat SOL density profile.

Based on the experimental plasma equilibrium for shot #103748 with flat SOL density profile, the pedestal density gradient is small and thus the pedestal pressure gradient and current are low. Thus, the pedestal is stable for ideal P-B mode with high pedestal collisionality cases and the linear growth rate is close to 0 for $v_{ped}^* = 0.415$ (0.8x), $v_{ped}^* =, 0.785$ (1.2x) and $v_{ped}^* = 1.316$ (1.2x), as shown by the red, blue and green curves in Fig. 12(a). When we further decrease the collisionality to $v_{ped}^* = 0.108$ (0.5x), the pedestal plasma becomes unstable with the most unstable mode number at high-n (n~70), possibly due to the large reduction of magnetic shear as shown by the black curve in Fig. 12(c). The high-n ballooning mode is marginally unstable with small growth rate ~0.012 as shown by the black curve in Fig. 13(a). The nonlinear simulations found that the fluctuation intensity increases, and a small ELM will be triggered when the collisionality decreases to $v_{ped}^* = 0.108$. Fig. 13(b) shows the spatial-temporal evolution of RMS of density perturbation vs normalized time and radius at the OMP with lower collisionality $v_{ped}^* = 0.108$ (0.5x case) and Fig. 13(c) shows the pressure profiles in the nonlinear phase for these four different collisionalities. The density fluctuation first grows up at outer midplane in the pedestal in the linear phase where the ballooning modes become unstable. When the fluctuation grows up to 6%, the pressure profile collapses, and then the pedestal falls into the linear stable zone just shortly after the initial ELM crash. Due to the marginal unstable and the limited turbulence spreading, there is no clear front penetration after initial ELM crash. The fluctuation intensity is still relative lower than that of the type-I ELM.

Finally, the flattening of pressure profile is localized only in a small radial area around the peak pressure gradient in the pedestal as shown by the black curve in Fig. 13(c), leading to a small ELM with ELM size <1%.

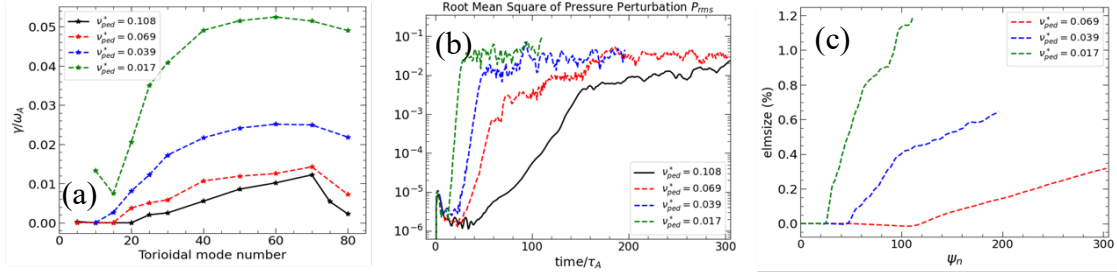

Fig. 14 (a) Linear growth rate from the linear toroidal mode number scan; (b) time evolution of RMS of pressure perturbation at peak gradient location at the OMP; (c) time evolution of 3D relative ELM size for shot #103748 with flat SOL density profile for low collisionality $v^*_{ped} \leq 0.108$.

As mentioned in the introduction, ITER will have even lower pedestal collisionality which is ~0.01. In order to project from the current tokamak to ITER relevant parameters, the collisionality is further decreased from 0.108 (0.5x case) to 0.069 (0.4x case), 0.039 (0.3x case) and 0.0107 (0.2x case) as shown in the last two rows of Table 3. The linear simulations show that the linear growth rates increase, and dominant mode number decreases as the collionality decreases, as show in Fig. 14(a). When the collisionality decreases to 0.0107, which is close to ITER parameters, the linear growth rate dramatically increases and the dominant mode shifts from n~70 to n~55-60 due to the increase of bootstrap current density and thus the enhanced peeling driven. Fig. 14(b) shows the contour plot of RMS value of pressure perturbation vs normalized time and radius at peak gradient location at the OMP in the nonlinear simulations. As collisionality decreases, the linear growth time reduces because of large growth rates and the saturated fluctuation level increases. The ELM size increases as the collisionaltiy decreases, as shown in Fig. 14(c). When the collisionality is decreased to 0.069 (0.4x case) and 0.039 (0.3x case), the turbulence spreading increases as the fluctuation intensity increases, but no front penetration is observed. The pressure profile collapse still remains in a small region inside the pedestal and the ELM size is saturated at a low level <1%, which are still in the small ELM regime, as shown by the dashed red and blue curves in Fig. 14(c). While for the lowest collisionality $v^*_{ped} = 0.0107$, the linear mode is very unstable with large linear growth rate and the pressure fluctuation intensity at the onset of nonlinear phase is stronger. The fluctuation intensity saturated in high level in nonlinear phase, leading to a strong inward avalanche with fast front propagation and deep penetration. Thus, the profile keeps collapsing after the initial ELM crash. The ELM size keeps increasing and a large ELM will be triggered in a short time scale driven by ideal peeling-ballooning mode with n~55-60, as shown by the dashed green curve in Fig. 14(c). When the collisionality decreases from 0.108 to 0.0107, the linear growth rate dramatically increases and the dominant mode shifts from n~70 to n~55-60 due to the increase of bootstrap current density and thus the enhanced peeling driver. When the collisionality decreases, two physical mechanisms can play a role. (1) magnetic shear reduces; (2) bootstrap current increases, leading to an increase of peeling driver. Both of them can lead to large growth rate. The dominant mode number shifted to low-n is due to the instability of peeling branch enhanced as a result of the bootstrap current increasing as collisionality decreases. Overall, the collisioanlity scan for the wide pedestal density profile (shot#103748) shows that small ELM can be triggered in a low collisionality regime with a window of collisionality $v_{ped}^*\sim 0.04-0.1$ and with a wide density pedestal, which is consistent with experimental observation[24,25].

**4.2 Pedestal density width/gradient scan**

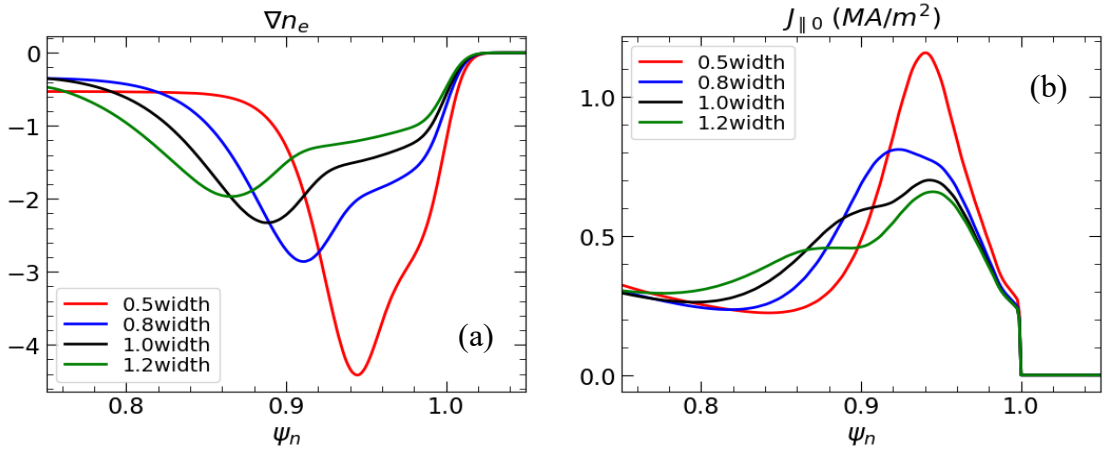

Fig. 15 (a) density gradient and (b) current density for pedestal density width scan.

To understand the impact of pedestal density gradient effect on the ELM dynamics, the scan of pedestal density profile width is performed using BOUT++ six-field two-fluid turbulence code. As mentioned in the introduction, a low pedestal density gradient is a key for access to the small/grassy ELM regime. ITER will operate in a low pedestal density gradient with low pedestal collisionality and high separatrix collisionality. Therefore, we start the scan from the 0.5x case for the collisionality scan of shot #103748 in section 4.2 (as the reference case), which is in a small ELM regime driven by marginal ballooning instability with low density gradient, and low collisionality $v_{ped}^* = 0.108$. Based on this reference case, the pedestal density width is scanned by multiplying a factor 0.5x, 0.8x, 1.0x and 1.2x with fixed temperature and thus the pressure profile changes accordingly. The pedestal density gradient increases with width decreases, as shown in Fig. 15(a). To self-consistently calculate the bootstrap current, the magnetic equilibria are reconstructed using the kinetic EFIT code with new density profiles. As the pedestal density width decreases, the pressure gradient increases, and thus the current density increases accordingly, as shown in Fig. 15(b).

The BOUT++ turbulence simulation results are shown in Fig. 16. Fig. 16(a) shows the linear growth rate vs. toroidal mode number n for ideal MHD instability with ion diamagnetic and Fig. 16(b) shows the time evolution of root mean square (RMS) value of pressure perturbation at peak gradient location at the OMP. The Black curve in the Fig. 16(a) is from the reference case, which is the black curve in Fig. 13(a) and 14(a). The small ELM is triggered by marginally unstable ballooning mode with high-n (n~70). When the width is increased by factor 1.2x, both the pedestal pressure gradient and current density decreases accordingly, and the high-n ballooning mode becomes stable. Thus, the linear growth rate decreases to 0, as shown by the green curve in Fig. 16(a). However, even though the linear simulations show that the growth rate is 0 for the ideal P-B mode, the nonlinear simulation shows that the growth rate in the linear phase is not 0 as shown by the green curve in Fig. 16(b), which is driven by the microturbulence instability, such as the drift-Alfvén insatiably. As reported by our previous paper[4], an ELM cannot be triggered by the drift-Alfvén insatiably even though it can drive a large linear growth rate. The pressure fluctuation saturates at a low level in the nonlinear phase and there is no ELM triggered with ELM size ~0 as shown by the green curve in Fig. 16(c). While when the pedestal density width decreases by factor 0.8x, the dominant unstable mode changes from high-n (n~70) to low-n (n~15) mode. Because the current density increases as the density gradient increases, the current driven instability enhances. Thus, the instability shifts from high-n ballooning mode to intermediate-n peeling-ballooning mode. But the maximum linear growth rate does not change much, as shown by the blue curve in Fig. 16(a). The pressure fluctuation saturates at the similar level to the reference case (black curve in Fig. 16(b)) in the nonlinear phase, leading to a small ELM driven by marginality peeling instability at the position of peak pressure gradient. The ELM size is comparable with the reference case as shown by the blue and black curve in Fig. 16(c). When the pedestal density width further decreases by factor 0.5x, the pedestal pressure gradient and current density increases dramatically. The pedestal is peeling-ballooning mode unstable with intermediate-n ~25. The linear growth rate increases dramatically, as shown by the red curve in Fig. 16(a). The pressure fluctuation saturates at a high level in the nonlinear phase, which is an order of magnitude higher than the small ELM. The mode peak shift inward and the pressure profile keeps collapsing inward to the pedestal top with a large turbulence spreading, leading to a large ELM burst. A large ELM is triggered by peeling-ballooning instability near the peak gradient of the pressure with a large ELM size >1%.

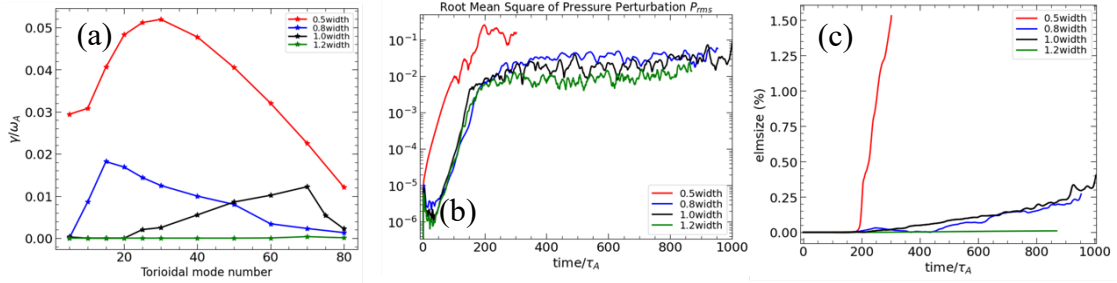

Fig. 16 (a) Linear growth rate vs toroidal mode number for ideal MHD instability with ion diamagnetic; (b) time evolution of root mean square (RMS) value of pressure perturbation at peak gradient location at the OMP; (c) time evolution of 3D ELM size for different pedestal density width.

Therefore, the pedestal collisionality scan and density width scan for shot # 103748 indicate that small ELM can be triggered by marginality ballooning instability and/or by marginality peeling instability with low pedestal collisionality $\nu_{ped}^*\sim 0.1$ for weaker pedestal density gradient.

## 5. Summary and discussion

Small ELMs are achieved via controlling strike points from vertical to horizontal divertor target from EAST experiments on the new lower tungsten divertor[14], which significantly modifies the pedestal density profiles. When the strike point is located on the vertical target (shot #103751), the separatrix density is low with a steep pedestal density gradient. Pedestal pressure gradient exceeds the ideal ballooning mode limit, large type-I ELM bursts are observed. When the strike point is shifted from the vertical target to horizontal target (shot#103745 and #103748), the separatrix density increases with weaker pedestal density gradients. Thus, both the pedestal density gradient and bootstrap current density reduce, small ELMs observed.

BOUT++ turbulence simulations are conducted to capture the physics of the small ELMs characteristics with different pedestal density profiles for EAST discharges with shot #103751, #103745 and #103748. As the pedestal density gradient decreases, BOUT ++ linear simulations show that the most unstable modes change from high-n ideal ballooning modes for shot #103751 to the intermediate-n peeling-ballooning modes for shot #103745, and the unstable modes are inside the pedestal region near the position of peak pressure gradient. The maximum growth rate is dramatically decreased due to the significant reduction of pedestal pressure gradient. When the pedestal density gradient further decreases, the pedestal plasma is eventually peeling-ballooning stable for shot #103748. However, a local instability near the separatrix is found driven by a large separatrix density gradient. Nonlinear simulations show that the fluctuation firstly grows up at the position of peak pressure gradient and finally saturated at a high level for the lowest SOL density case with shot #103751. The pressure profile keeps collapsing after the initial ELM crash, leading to a large ELM. As the pedestal density gradient decreases due to the change of the strike point from vertical to horizontal divertor target from EAST experiments, the saturated pressure fluctuation level decreases and the ELM size decreases, which becomes less than 1%, leading to small ELMs. Simulations further show that the local instability grows up near the separatrix with a larger separatrix density gradient for a stable pedestal plasma with shot #103748, which also can trigger small ELMs and enhance the SOL turbulence transport. Therefore, based on pedestal plasmas from three EAST discharges, BOUT++ turbulence simulations show that small ELMs can be triggered either by the marginally ideal peeling-ballooning instabilities near the peak pressure gradient position inside the pedestal or local instabilities in the pedestal foot with a large separatrix density gradient.

The pedestal collisionality and density gradient play an important role in the pedestal MHD stability for access to the small ELM regime. As we mentioned in section 4.1, ITER probably will operate in the low pedestal collisionality regime with low pedestal density gradient[15]. Here is a question. Will ITER be able to access the small ELM regime with wide pedestal density but low pedestal collisionality? To find out the window of collisionality for access to small ELM regime, a scan of pedestal collisionality is performed using BOUT++ turbulence code by decreasing density and increasing temperature with a fixed pressure profile. Pedestal collisionality scan for shot #103751 with steep pedestal density gradient and type-I ELMs shows the pedestal is unstable for high-n ideal ballooning modes. The linear growth rate decreases and ELM size decreases with increasing pedestal collisonality. The simulation results for the ELM size reduction with increasing pedestal plasma collisionality for EAST type-I ELM show a good agreement with the ITPA multi-tokamak experimental database and previous BOUT++ simulation results. Pedestal collisionality scan for EAST shot #103748 with weak pedestal density gradient and small ELMs shows that as the pedestal collisionality decreases, pedestal plasmas undergo a transition from stable to unstable to ideal peeling-ballooning modes, yielding a window of collisionality ( $0.04 < \nu_{ped}^* < 0.1$) for small ELMs. When the collisionality further decreases, the ELM size increases, which leads to large ELM when the collisionality reduces to $\nu_{ped}^* \sim 0.01$. Therefore, small ELMs can be triggered in the low collisionality regime $0.04 < \nu_{ped}^* < 0.1$ with weak pedestal density gradient. The weak pedestal density gradient is a key for access to the small ELM regime. The pedestal density width scan from small ELMs driven by marginally unstable ballooning modes with low collisionaltiy $\nu_{ped}^* \sim 0.1$ shows that the dominated mode shifts from high-n ballooning mode to low-n peeling mode when the pedestal width decreases, and eventually to intermediate-n peeling-ballooning mode, leading to large type-I ELMs with narrow pedestal width. Therefore, from the simulations, we show that small ELM regime can be achieved in a low normalized pedestal collisionality regime with wide pedestal density / weak pedestal density gradient expected in ITER. However, the simulations are based on the EAST equilibria and other important control parameters are still far away with the ITER scenarios, such as $q_{95}$, $\beta_p$, $\delta$ etc. Further simulations based on ITER scenarios are needed to investigate the access to the small ELM regime.

Overall, the BOUT++ simulations indicate that the small ELM can be triggered by marginally unstable mode with different type mode, such as low-n peeling mode, high-n ballooning mode, intermediate-n peeling-ballooning mode or a local instability in the pedestal foot with a larger separatrix density gradient. The ELM small or large strongly depends on the inward avalanche or turbulence spreading which includes the front propagation and penetration. For the large ELM, the linear mode is very unstable with large linear growth rate and the pressure fluctuation intensity at the onset of nonlinear phase is much strong, leading to an ELM crashing. After the initial ELM crashing, the original pedestal profile collapses, which results in the profile steepening inward near the pedestal top. The pedestal top gets into the linear unstable zone, which leads to a $2^{nd}$ collapses. These processes continue to generate multiple crashing inward until the pedestal gets into the linear stable zone. The multiple profiles collapse from linear unstable zone near original pedestal peak gradient position into stable zone in the pedestal top region, which tap more pedestal stored energy into fluctuation energy, leading to large ELMs. This process is termed as avalanche process. The front propagation follows the sequence of multiple profiles collapsing. While for small ELMs, the pedestal is near linear instability threshold and pressure fluctuation intensity at the onset of nonlinear phase is much weaker, the fluctuation is limited near the peak gradient position of the pedestal with only diffusive penetration, but without the front propagation. Further research is needed to determine the scaling of the speed of fluctuation intensity front propagation and the penetration depth into linear stable zone by turbulence spreading.

## Acknowledgement

The authors would thank Dr. Ze-Yu Li, Huiqian Wang, Qingquan Yang and Ben Zhu for useful discussions. This work was performed under the U.S. Department of Energy by Lawrence Livermore National Laboratory under Contract No. DE-AC52-07NA27344, LLNL-JRNL-837352. This work was also supported by the Users with Excellence Program of Hefei Science Center, CAS under grant Nos. 2021HSC-UE014 and by the National Key R&D Program of China Nos. 2017YFE0301206, 2017YFE0300402 and 2017YFE0301100.

# AUTHOR DECLARATIONS
## Conflict of Interest

The authors have no conflicts to disclose.

## DATA AVAILABILITY

The data that support the findings of this study are available from the corresponding author upon reasonable request.

# References

[1] R.J. Goldston, Nucl. Fusion 52 (2012) 013009.

[2] T. Eich, B. Sieglin, A. Scarabosio, W. Fundamenski, R. J. Goldston, A. Herrmann and ASDEX Upgrade Team, Phys. Rev. Lett. 107 (2011) 215001.

[3] A. Loarte, B. Lipschultz, A.S. Kukushkin, G.F. Matthews, P.C. Stangeby, N. Asakura, G.F. Counsell, G. Federici, A. Kallenbach, K. Krieger, et al., Nuclear Fusion, 47 (2007): S203-S263.

[4] N.M. Li, X.Q. Xu, Y.F. Wang, N. Yan, J.Y. Zhang, J.P. Qian J. Z. Sun and D.Z. Wang, Nucl. Fusion 62 (2022) 096030.

[5] T.F. Tang, X.Q. Xu, G.Q. Li, J.L. Chen, V.S. Chan, T.Y. Xia, X. Gao, D.Z. Wang and J.G. Li, Nucl. Fusion 62 (2022) 016008.

[6] X.Q. Xu, N.M. Li, Z.Y. Li, B. Chen, T.Y. Xia, T.F. Tang, B. Zhu and V.S. Chan, Nucl. Fusion 59 (2019)126039.

[7] Ze-Yu Li, X.Q. Xu, N.M. Li, V.S. Chan and X.G. Wang, Nucl. Fusion 59 (2019) 046014.

[8] R. Perillo, J.A. Boedo, C.J. Lasnier, I. Bykov, C. Marini, and J.G. Watkins, Phys. Plasmas 29 (2022) 052506.

[9] G.S. Xu, Q.Q. Yang, N. Yan, Y.F. Wang, X.Q. Xu, H.Y. Guo, R. Maingi, L. Wang, J.P. Qian, X.Z. Gong, V.S. Chan et al., Phys. Rev. Lett. 122 (2019) 255001.

[10] R. Nazikian, C.C. Petty, A. Bortolon, X. Chen, D. Eldon, T.E. Evans, B.A. Grierson, N.M. Ferraro, S.R. Haskey, M. Knolker et al., Nucl. Fusion 58 (2018) 106010.

[11] X.Q. Xu, "Divertor Heat Flux Broadening by Grassy ELMs", 28th IAEA Fusion energy conference (FEC 2020), https://conferences.iaea.org/event/214/contributions/17601/.

[12] B. Labit, T. Eich, G.F. Harrer, E. Wolfrum, M. Bernert, M.G. Dunne, L. Frassinetti, P. Hennequin, R. Maurizio, A. Merle, et al., Nucl. Fusion 59 (2019) 086020.

[13] J. Garcia, E. de la Luna, M. Sertoli, F.J. Casson, S. Mazzi, Z. Stancar, G. Szepesi, D. Frigione, L. Garzotti, F. Rimini et al., Phys. Plasmas 29 (2022) 032505.

[14] G.S. Xu, X. Lin, Q.Q. Yang, Y.F. Wang, G.Z. Jia, N.M. Li, N. Yan, R. Chen, X.Q. Xu, H.Y. Guo et al., "Small-ELM-regime access facilitated by new tungsten divertor on EAST", 48th EPS Conference on Plasma Physics,
https://indico.fusenet.eu/event/28/contributions/264/attachments/278/635/EPSpaper2022-Xu.pdf.

[15] A. Loarte, "Required R&D in Existing Fusion Facilities to Support the ITER Research Plan", ITR-20-008, 22 September 2020,
https://www.iter.org/doc/www/content/com/Lists/ITER%20Technical%20Reports/Attachments/14/ITR_20_008_Required_RD_in_existing_fusion_facilities_to_support_the_ITER_Research_Plan.pdf.

[16] L.L. Lao H. St. John, R.D. Stambaugh, A.G. Kellman and W. Pfeiffer, Nucl. Fusion 25 (1985)1611.

[17] O. Meneghini, S.P. Smith, L.L. Lao, O. Izacard, Q. Ren, J.M. Park, J. Candy, Z. Wang, C.J. Luna, V.A. Izzo, et al., Nuclear Fusion, 55 (2015) 083008.

[18] O. Meneghini, P.B. Snyder, S.P. Smith, J. Candy, G.M. Staebler, E.A. Belli, L.L. Lao, J.M. Park, D.L. Green, W. Elwasif, B.A. Grierson and C. Holland, Physics of Plasmas, 23(2016) 042607.

[19] R.J. Groebner, T.H. Osborne, Physics of Plasmas, 1998, 5(5): 1800-1806

[20] T.Y. Xia, X.Q. Xu and P.W. Xi,. Nuclear Fusion, 53(2013) 073009.

[21] B. Zhu, H. Seto, X.Q Xu, and M. Yagi, Computer Physics Communications, 267 (2021) 108079.

[22] A. Loarte, G. Saibene, R. Sartori, D. Campbell, M. Becoulet, L. Horton, T. Eich, A. Herrmann, G. Matthews, N Asakura et al., Plasma Phys. Control. Fusion 45 (2003) 1549–1569.

[23] X.Q. Xu, J.F. Ma and G.Q. Li, Physics of Plasma 21(2014) 120704.

[24] N. Oyama, P. Gohil, L.D. Horton, A.E. Hubbard, J.W. Hughes, Y. Kamada, K. Kamiya, A.W. Leonard, A. Loarte, R. Maingi et al., Plasma Physics and Controlled Fusion 48 (2006): A171-A181.

[25] Y.R. Zhu, Ze-Yu Li, V.S. Chan, J.L. Chen, X. Jian, B.D. Dudson, A.M. Garofalo, P.B. Snyder, X.Q. Xu, G. Zhuang and CFETR physics team, Nuclear Fusion 60 (2020) 046014.